\newcommand{\CLASSINPUTtoptextmargin}{2cm}
\documentclass[conference]{IEEEtran}

\IEEEoverridecommandlockouts

\ifCLASSINFOpdf

\else

\fi

\usepackage{textcomp}
\usepackage{graphicx}
\usepackage{float}
\usepackage{cite}
\usepackage{booktabs}
\usepackage{epstopdf}
\usepackage{multirow}
\usepackage{boldline}
\usepackage{amsmath}
\usepackage{relsize}
\usepackage{array}
\usepackage{makecell}
\usepackage{colortbl}
\usepackage{algorithm}
\usepackage{algpseudocode}
\usepackage{soul}
\usepackage[normalem]{ulem}
\usepackage{flushend}

\usepackage[normalem]{ulem}
\newcommand\soo{\bgroup\markoverwith{\textcolor{red}{\rule[0.5ex]{2pt}{0.4pt}}}\ULon}

\newcommand{\ve}[1]{\boldsymbol{#1}}

\def\infinity{\rotatebox{90}{8}}

\makeatletter
\newlength \figwidth
\if@twocolumn
\else
\fi
\makeatother
\makeatletter
\def\ps@IEEEtitlepagestyle{%
  \def\@oddfoot{\mycopyrightnotice}%
  \def\@evenfoot{}%
}
\def\mycopyrightnotice{%
  {\footnotesize \textit{This article is accepted at the IEEE International Conference on Communications (ICC)} $\mathit{2018}$\hfill}
  \gdef\mycopyrightnotice{}
}

\begin{document}
\IEEEoverridecommandlockouts
%
\title{Cross-Layer Designs for Body-to-Body Networks: Adaptive CSMA/CA with Distributed Routing}
\author{\IEEEauthorblockN{Samiya~M.~Shimly}
\IEEEauthorblockA{The Australian National University$^\mathsection$\\
CSIRO Data61$^\dagger$\\
Email: Samiya.Shimly@data61.csiro.au}\vspace{-20pt}
\and
\IEEEauthorblockN{David~B.~Smith}
\IEEEauthorblockA{CSIRO Data61$^\dagger$\\
The Australian National University\\
David.Smith@data61.csiro.au}\vspace{-20pt}
\and
\IEEEauthorblockN{Samaneh~Movassaghi}
\IEEEauthorblockA{The Australian National University$^\mathsection$\\
CSIRO Data61$^\dagger$\\
Samaneh.Movassaghi@data61.csiro.au}\vspace{-20pt}
\thanks{$^\dagger$National ICT Australia (NICTA) has been incorporated into Data61 of CSIRO. $^\mathsection$This research is supported by an Australian Government Research Training Program (RTP) scholarship.}}

\maketitle
\begin{abstract}
In this paper, we propose a novel adaptive carrier sense multiple access scheme with collision avoidance (CSMA/CA) to perform efficient and reliable data transfer with increased throughput across multiple coexisting wireless body area networks (BANs) in a tiered architecture. We investigate the proposed scheme using two distributed cross-layer optimized dynamic routing techniques, i.e., shortest path routing (SPR) and cooperative multi-path routing (CMR). The channel state information from the physical layer is passed on to the network layer using an adaptive cross-layer carrier sensing mechanism between the physical and MAC layer, which adjusts the carrier sense threshold (e.g., RSSI) periodically based on the slowly-varying channel condition. An open-access experimental dataset of `everyday' mixed-activities is used for analyzing the  cross-layer optimization. Our proposed optimization using adaptive carrier sensing performs better than static carrier sensing with CSMA/CA as it reduces the continuous back-off duration and latency as well as significantly increases the throughput (in successful packets/s) by more than 50\%. Adaptive CSMA/CA also shows 20\% and 6\% improvement over a coordinated TDMA approach with higher duty cycle for throughput and spectral efficiency, respectively, and provides acceptable packet delivery ratio and outage probability with respect to SINR.
\end{abstract}

\IEEEpeerreviewmaketitle 
\section{Introduction}
Wireless Body Area Networks (BANs) are an exciting networking technology for body-centric wireless communications, where low power, short-range micro and nano-technology sensors/actuators are placed on, in, around or/and near the human body to monitor physiological signals. Aside from having a diverse range of applications, these networks are principally focused on advanced healthcare management, specifically in medical rehabilitation, diagnosis and monitoring patients. When multiple closely-located BANs coexist, the potential inter-network communication and cooperation across BANs leads to the implementation of wireless body-to-body networks (BBNs) \cite{meharouech2015future}. The main motivation behind BBNs is to make use of body-to-body communication to overcome the problems of coexistence and general performance degradation for closely located BANs. This type of network could provide cost-effective solutions for remote monitoring of a group of patients in emergency incidents, for instance, by relaying physiological data from body-to-body up to the access point of the medical server in case of out-of-range network infrastructure. The notion of BBN is more dynamic and potentially large-scale, where each BAN can join and/or leave the network seamlessly, without the need for any centralized infrastructure. Besides, due to the highly mobile nature of BANs, it is generally not feasible to assign a central coordinator among BANs to maintain coexistence \cite{dong2013opportunistic}.

Along with some centralized multiple access mechanisms \cite{cheng2013coloring, dong2013opportunistic, movassaghi2016enabling} for radio interference mitigation in coexisting BANs, few recent works \cite{grassi2012b2irs, chen20142l, huang2015adaptive} consider carrier sense multiple access with collision avoidance (CSMA/CA) techniques to deal with network dynamics and mobility of BANs. A MAC protocol with two-layer interference mitigation (2L-MAC) is proposed in \cite{chen20142l}, where carrier sensing and polling frames are used to reduce inter-BAN and intra-BAN interference, respectively. To make the approach in \cite{chen20142l} more flexible, the authors in \cite{huang2015adaptive} proposed an adaptive CSMA/CA MAC protocol for BANs at the coordinator level, which adjusts the polling period/MAC frame length according to the interference level. However, all of this recent work with CSMA/CA only focuses on intra-BAN transmissions when considering inter-BAN interference.

A back-off-tuning algorithm for CSMA/CA was proposed in \cite{cali2000ieee} for the IEEE 802.11 Standard to increase the successful transmissions by controlling the back-off counter, which does not consider the effect of latency and packet failure rate on the channel. In \cite{ho2007impact}, the authors used adaptive distance-based transmit power control with CSMA/CA to increase throughput which is not suitable for power-constrained body area networks. In recent work \cite{chau2017effective}, the authors proposed an adaptive CSMA/CA mechanism considering the IEEE $802.11$ Standard (WiFi networks) for cumulative-interference-power carrier-sensing (CPCS) and pairwise-distance based incremental-power carrier-sensing (IPCS), where the carrier sense thresholds are adjusted instantly based on the dynamic feedback of nearby transmissions, with respect to the detected level of hidden nodes and exposed nodes. Yet, this approach consumes more energy due to a frequent change of the carrier sensing threshold, which is not efficient for slowly-varying and extremely resource-constrained networks like BANs that have more stringent requirements than other standards subject to transmit power, mobility and topological changes \cite{movassaghi2016enabling}. 

\vspace{0.5mm}In this paper, we propose an adaptive cross-layer CSMA/CA scheme with maximum-interference-power carrier-sensing (MPCS) \emph{to enable efficient communication amongst multiple coexisting BANs, adjusting the carrier sense threshold based on the slowly-varying channel conditions with an adequate periodic time-stamping, to obtain a suitable trade-off between the amount of interference, throughput, latency and energy consumption of CSMA/CA channels.} We analyze the performance of the proposed adaptive CSMA/CA by performing decentralized cross-layer optimization across the physical, MAC and network layers for two-tiered communications, with on-body BANs at the lower tier and a BBN at the upper tier. The analysis is applied to an open-access radio measurement dataset provided in \cite{smith2012body} (captured using NICTA$^\dagger$ developed wearable channel sounders/radios), recorded from `everyday' mixed-activities and a range of measurement scenarios with people wearing radios. From the implementation of the proposed adaptive CSMA/CA with the experimental measurements used in this paper, we demonstrate the following:
\begin{itemize}
    \item With the proposed adaptive carrier sensing, the percentage of longer (greater than $3$~s) continuous back-off duration is trivial ($0$\%), whereas with static threshold (e.g., $-95$ dBm), the channel can remain continuously in back-off for very long period (more than $1000$~s) with $6$\% of the total time.
    \item The average continuous back-off duration of the CSMA/CA channels with the proposed adaptive carrier sensing mechanism is $237$ ms, whereas it can goes up to $11$~s on average with a static carrier sensing threshold of $-95$ dBm.
     \item The proposed adaptive mechanism can provide more than $50\%$ improvement over static carrier sensing, in terms of throughput (successful packets/s) and packet arrival rate.
     \item Even though static carrier sensing provides better outage probability with respect to signal-to-interference-plus-noise-ratio (SINR) as it is strict towards avoiding larger interference levels, it suffers from decreased throughput and increased latency.
     \item Adaptive CSMA/CA provides up to $20\%$ and $30\%$ improvement over TDMA (with a higher duty cycle of $8.3\%$), in terms of average throughput (successful packets/s) and packet arrival rate, respectively.
    \item For spectral efficiency, adaptive CSMA/CA provides up to $6\%$ improvement over TDMA with $8.3\%$ duty cycle.
\end{itemize}

\vspace{0.1em}The rest of this paper is organized as follows: Section II presents the system model and the experimental scenario. In Section III, the proposed adaptive carrier sensing mechanism is described along with a brief demonstration of the applied routing techniques. The performance of adaptive CSMA/CA is analyzed and compared with static CSMA/CA and TDMA in Section IV. Finally, concluding remarks are provided in Section V.

\section{System Model}
\begin{figure}[!t]
\centering{\includegraphics[width=0.9\figwidth]{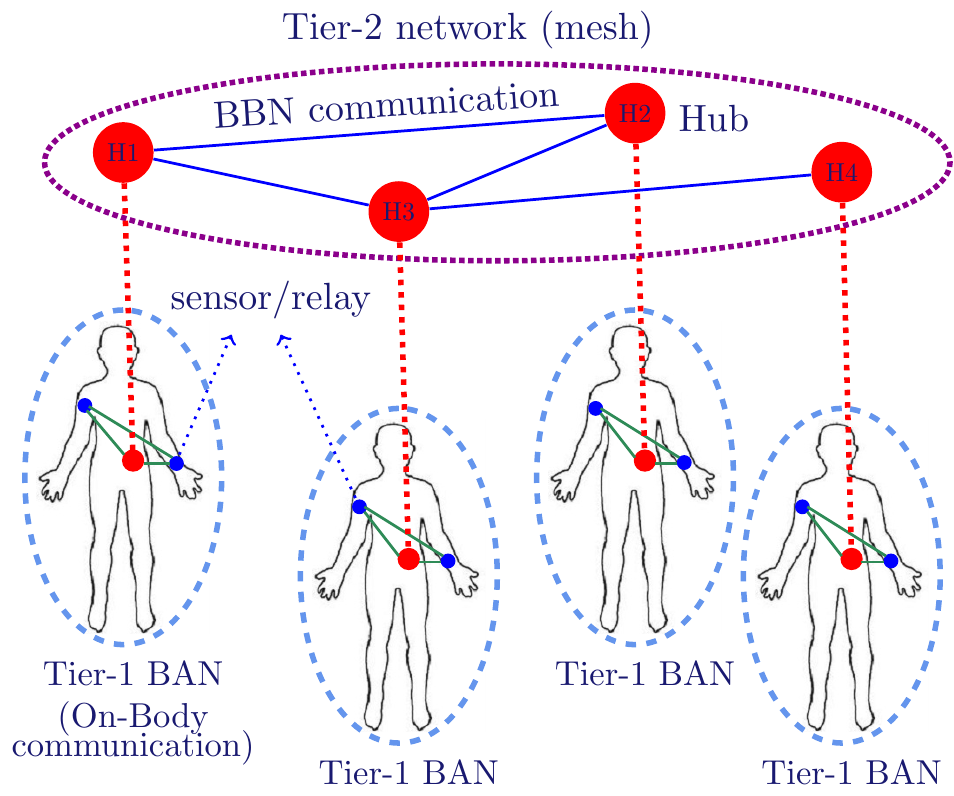}}
\caption{Two tiered architecture with $4$ colocated BANs; Hub on the left-hip and two sensors/relays on the left-wrist and right-upper-arm, respectively.}
\label{tier}
\end{figure}

In this work, $10$ co-located mobile BANs (people with fitted wearable radios) deployed for experimental measurements are considered in a decentralized two tiered architecture, where the hubs of the BANs are deployed in a mesh topology in the upper tier (for inter-BAN/BBN communication) and the on-body sensors/relays of the corresponding BANs are deployed in the lower-tier (for intra-BAN communication). It can be portrayed as a hybrid mesh architecture where BANs (hubs/gateways) are performing as both clients and routers/relays, which will enable flexible and fast deployment of BANs to provide greater radio coverage, scalability and mobility. An illustration of the two tiered architecture with $4$ coexisting BANs (with fitted on-body hubs and sensors) is given in Fig.\ref{tier}.
\setlength{\textfloatsep}{2pt}
Dynamic routing is performed at the network layer in a cross-layered approach using the authors proposed routing techniques, i.e., SPR and CMR \cite{shimly2017cross}, that utilize and interact with the physical layer. Therefore, changes in channel states detected at the physical layer are passed on to the network layer, so routes with the most favorable channel conditions are chosen.

\vspace{-1mm}\subsection{Experimental scenario}
The open-access dataset used in our analysis, consists of continuous extensive intra-BAN (on-body) and inter-BAN (body-to-body) channel gain data of around $45$ minutes, captured from $10$ closely located mobile subjects (adult male and female). We consider $10$ mobile people as a reasonable amount for BAN coexistence experimentation, due to the dramatic impact caused by the slowly-varying human-body dynamics and shadowing by body-parts, both on the on-body and inter-body channels \cite{smith2015channel}. Also this is an acceptable amount of coexisting BANs to be supported by the physical layer according to the IEEE $802.15.6$ Standard \cite{tg6_std}.

The experimented subjects were walking together to a hotel bar, sitting there for a while and then walking back to the office. Each subject wore $1$ transmitter (hub) on the left-hip and $2$ receivers (sensors/relays) on the left-wrist and right-upper-arm, respectively. The radios were transmitting at $0$ dBm transmit power with $-100$ dBm receive sensitivity. A description of these wearable radios and hardware platform can be found in \cite{hanlen2010open} and the experimental dataset can be downloaded from \cite{smith2012body}. Each transmitter transmits in a round-robin fashion, at $2.36$ GHz, with $5$ ms separation between each other. Hence, each transmitter is transmitting in every $50$ ms (with a sampling rate of $20$ Hz) to all $9$ other subject's receivers as well as their own receivers (all small body-worn radios/hubs/sensors), along with capturing the RSSI (Receive Signal Strength Indicator) values in dBm\footnote{Any received signal strength below $-$100~dBm, resulting in incorrectly decoded packets in experiment, is set at $-$101~dBm in analysis}, which gives a total of $300$ channel measurements (including both on-body and body-to-body links) in real-time over the whole network during the measurement period. The samples of given links are concatenated contiguously over each $600$ ms timestamp period. Due to the reciprocity property, the channel from any $T_x$ (transmitter) at position $a$ to $R_x$ (receiver) at position $b$ is similar for $T_x$ at $b$ to $R_x$ at $a$ \cite{hanlen2010open}, thus transmitters and receivers can be considered interchangeably.

\section{Proposed Adaptive carrier sensing mechanism}
Adaptive CSMA/CA with maximum-interference-power carrier-sensing (MPCS) is employed as the co-channel access scheme across (inter-BAN/coordinator level)  and within (intra-BAN/sensor level) all BANs to enable co-channel interference mitigation without global coordination. For effective dynamic estimation of the channel, continuous time-stamping is used in each $600$ ms which is reasonable, given the longer coherence times of $500$ ms (up to $1$ s) for `everyday' activities of narrowband on-body BAN channels \cite{smith2013propagation}, and for body-to-body channels used here where we have calculated the coherence time to be $900$ ms (with autocorrelation coefficient of $0.7$).

Whenever a node is ready to transmit data, it checks the availability of the channel by measuring the maximum interference power from the potential interference caused by the surrounding nodes that are trying to access the channel at the same time. For a given time instant $\tau$ of $i^{th}$ timestamp ($\tau_i$), the transmission of a given link (signal-of-interest from source to destination) is permitted by the simple carrier sensing mechanism with collision avoidance by MPCS, if
\begin{align}\label{csma/ca}
\max(\ve{\eta})&< cs_{th_i}\\
\textrm{where } \ve{\eta}&=\big(p_{tx} \left | h_{int_k,d}(\tau_i)\right | ^2\big),\,k=1,\ldots,n\nonumber
\end{align}
where $p_{tx}$ is the transmit power, $\left | h_{int_k,d}(\tau_i)\right |$ is the interfering channel gain from the $k^{th}$ interfering BAN to the destination $d$ at time instant $\tau$ of $i^{th}$ timestamp, $n$ is the number of interfering BANs and $cs_{th_i}$ is the adaptive carrier sense threshold (e.g., RSSI) in $i^{th}$ timestamp. If the condition in (\ref{csma/ca}) is not fulfilled, the node defers its transmission for a back-off period (still sensing the medium) until it finds the channel available for transmission. When the channel is found available by the node, it waits for a short inter-frame space period ($T_{SIFS}$) and then transmits the data. If an acknowledgement (ACK) is not received by the node, it implies a failure has occurred and the node tries to retransmit the data with the same procedure.

We apply adaptive carrier sensing to reduce the longer back-off period, hence improving throughput and end-to-end delay of the overall network. The channel conditions (e.g., back-off percentage, ill-conditioned periods of incorrectly decoded packets) of each timestamp are used as an approximation of the channel conditions of the next timestamp:
\begin{equation}\label{csth}
X_{i+1} \approx X_i, \quad i\geq 1
\end{equation}

\begin{algorithm}[t!]
\caption{Estimating Adaptive Carrier Sense Threshold}\label{alg_adp}
\begin{algorithmic}[1]
\State $CS_{th_i} \gets \textit{carrier sense threshold for the $i^{th}$ timestamp}$
\State $TS_i \gets \textit{outcome of CSMA/CA for the $i^{th}$ timestamp}$
\State $BOP_i \gets \textit{estimate back-off percentage of } TS_i $
\State $ICP_i \gets \textit{estimate the percentage of ill-conditioned}$
\State \hspace{12mm} $\textit{period of } TS_i$
\If {$BOP_i >= 50\%$}
\State $CS_{th_{i+1}} \gets CS_{th_i} + 1$
\Else
\If {$BOP_i < 50\%$ \& $ICP_i > 50\%$}
\State $CS_{th_{i+1}} \gets CS_{th_i} - 1$
\Else
\State $CS_{th_{i+1}} \gets CS_{th_i}$
\EndIf
\EndIf
\end{algorithmic}
\end{algorithm}

where for timestamp $i$, the channel condition of the next timestamp $X_{i+1}$ can be estimated from the current channel condition $X_i$. This estimation is a systematic approach for dynamic prediction of the inter-BAN channels given the longer coherence time of around $900$ ms and the timestamp period ($600$ ms) used in this work. In our proposed method, if the probability of back-off period is higher than 50\%, the carrier-sense threshold is adjusted to permit more transmissions, despite the amount of interference, to reduce continuous latency. Additionally, if the probability of transmitting incorrectly decoded packets is higher than 50\% even though having a lower back-off duration, the threshold is adjusted to decrease the amount of interference to reduce packet failure rate. This way there is a suitable trade-off between the latency, throughput and amount of interference of the CSMA/CA channels. The RSSI threshold for the first timestamp is predicted to be $-90$ dBm from an estimated median of typical on-body and inter-body RSSI measurements. The dynamic setting of the carrier-sense threshold is described in Algorithm \ref{alg_adp}.

We use two cross-layer optimized dynamic routing techniques (proposed in \cite{shimly2017cross}) to analyze the performance of adaptive CSMA/CA, which are implemented based on the Open Shortest Path First (OSPF) protocol using Dijkstra's algorithm. The routing table is updated in each timestamp and data is routed dynamically across the chosen path, determined based on the channel state information (i.e., ETX, hop count) of the previous timestamp. The routes for the first timestamp are estimated with a randomly selected value ($\geq 1$ and $< \infinity$) for each link, based on the defined routing metric, i.e., ETX.

\subsubsection{Dynamic Shortest path routing}
In our proposed method for shortest path routing, we used two different routing metrics: Expected Transmission Count (ETX) and hop count. The ETX metric can be calculated as follows:
\begin{equation}\label{etx}
ETX = (1-O_p)^{-1}
\end{equation}
where $O_p$ is the outage probability.
We chose an optimal path with lowest cost possible by combining these two metrics (ETX + Hop count) with a restriction to two hop counts, providing a trade-off between reliability and energy consumption.

\subsubsection{Dynamic Cooperative multi-path routing}
We use a novel Cooperative Multi-path Routing (CMR) scheme that we proposed and described in \cite{shimly2017cross}, which uses multiple routes between source and destination and employs $3$-branch selection combining (SC) within individual route paths. This method incorporates shortest path routing (SPR) and improves the performance of SPR by exploiting the benefits of multi-path diversity and shared resources \cite{quer2013inter}. The process for CMR is summarized in Algorithm \ref{alg_cmr}.
\begin{algorithm}[t!]
\caption{Estimating output of CMR, incorporating SPR (with ETX + max. $2$ hops count)}\label{alg_cmr}
\begin{algorithmic}[1]
\State $\{S,D\} \gets \textit{\{Source node, Destination node\}}$
\Function{FindShortestPath}{$S$,$D$}
\State $P_{etx} \gets \textit{ETX values of all possible paths from S to D}$
\State $[i , j] \gets [1 , \textit{size of } P_{etx}]$
\While{$i\ne j$}
\State $temp \gets \textit{Find min }(P_{etx})$
\If{$hop\_count(temp) = 2$}
\State $S\_to\_D \gets Path(temp)$
\Else
\State $P_{etx} \gets P_{etx} - \textit{temp}$
\State $temp \gets \textit{Find min }(P_{etx})$
\State $j \gets j - 1$
\EndIf
\State $i \gets i + 1$
\EndWhile
\If {$S\_to\_D$ is empty}
\State $S\_to\_D \gets \textit{direct path}$
\EndIf
\State \textbf{return} $S\_to\_D$
\EndFunction
\State $P_1 \gets FindShortestPath(S,D)$
\State $P_2 \gets \textit{FindShortestPath(S, D)} \notin P1$
\For {$i \gets 1,2$}
\State $P_1\_RH_i \gets selection\_combining(route\_hop_i)$
\EndFor
\State $Comb\_P_1 \gets min(P_1\_RH_1, P_1\_RH_2)$
\State $Comb\_P_2 \gets \textit{Repeat steps $24$ to $27$ for $P_2$}$
\State $\textbf{Output\_CMR} \gets max(Comb\_P_1, Comb\_P_2)$\end{algorithmic}
\end{algorithm}
\section{Performance Analysis}
In this section, we discuss and compare the results found from adaptive CSMA/CA applied on SPR and CMR techniques with the experimental measurements.  We compare the performance of applying static and adaptive carrier sensing thresholds for CSMA/CA in case of continuous back-off duration and throughput (successful packets/s) vs. packet arrival rate and outage probability with respect to SINR. We also provide a performance comparison between adaptive CSMA/CA and coordinated TDMA in terms of throughput vs. packet arrival rate, packet delivery ratio and spectral efficiency. The applied parameters for the performance analysis are listed in Table \ref{table_param}.
\begin{table}[!h]
\centering
\caption{Applied Parameters}
\label{table_param}
\begin{tabular}{|c|c|}\Xhline{0.8pt}
Parameter & Value\\[1ex]\Xhline{1pt}
Bandwidth ($B$) & $1$ MHz\\\hline
Carrier Frequency & $2.36$ GHz\\\hline
Data rate & $486$ kbps\\\hline
Packet size ($\ell$) & $273$ bits\\\hline
$T_{SIFS}$ & $50$ $\mu s$\\\hline
Packet transmission time ($T_{packet}$) & $0.6$ ms\\\hline
$T_{ACK}$ & $0.2$ ms\\\hline
Transmit power ($p_{tx}$) & $0$ dBm\\\hline
Total time ($T$) & $45$ mins\\\Xhline{0.8pt}
\end{tabular}
\end{table}

\vspace{-1mm}\subsection{Continuous Back-off Duration}
\begin{figure}[!t]
\centering{\includegraphics[width=\figwidth]{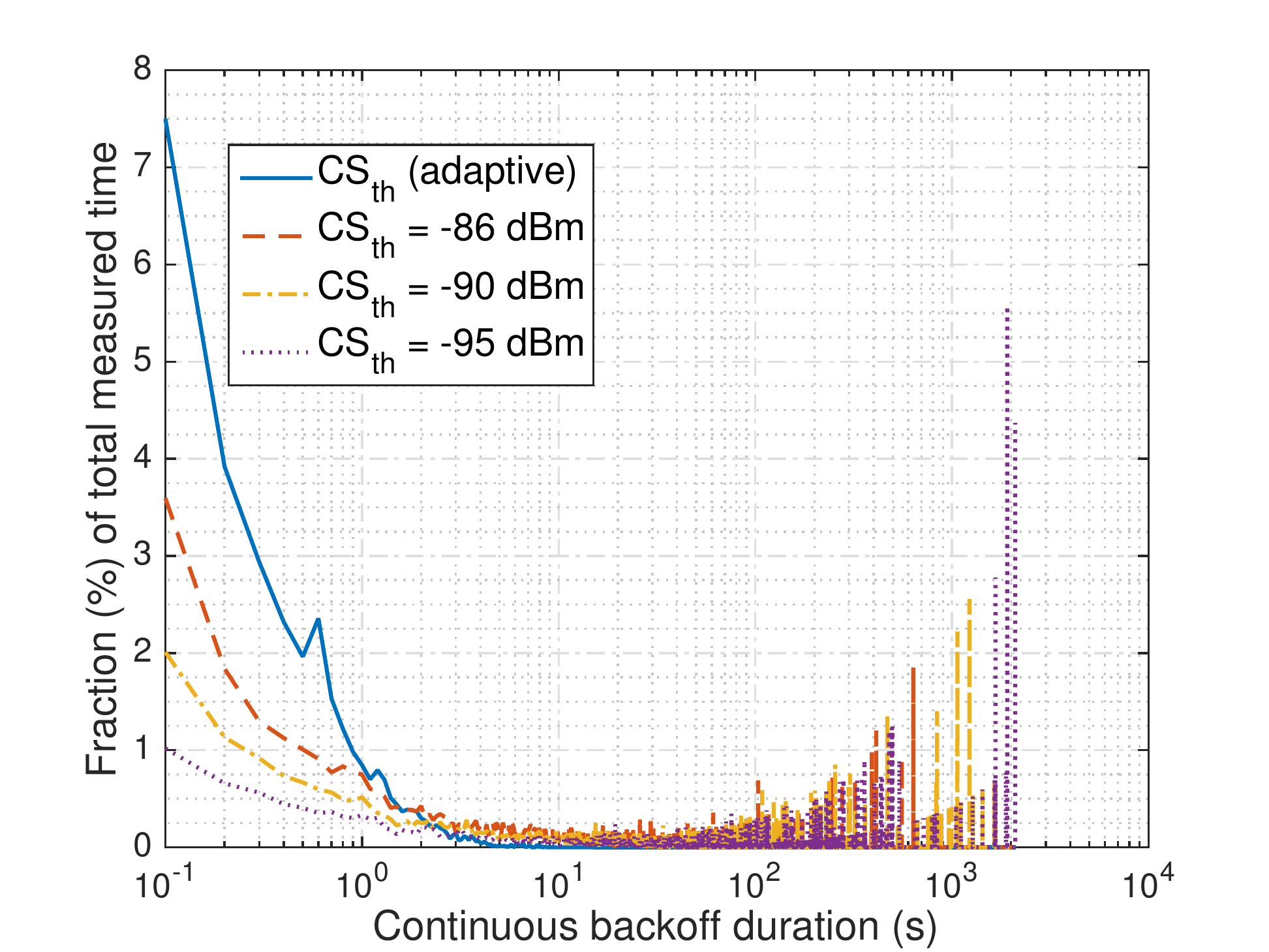}}
\caption{Percentage of continuous back-off duration of CSMA/CA links with an adaptive carrier sense threshold ($CS_{th}$) and different static carrier sense thresholds ($CS_{th}$), at transmit power $0$ dBm, with $10$ coexisting BANs.}
\label{bp}
\end{figure}

Since the continuous back-off duration is a key contributor to latency, this duration is an important performance metric for CSMA/CA channels. The continuous back-off durations of the CSMA/CA channels for body-to-body communications with different static carrier sense thresholds and an adaptive threshold, applied according to Algorithm \ref{alg_adp}, are shown in Fig. \ref{bp}.
It is demonstrated that apart from having a higher percentage of shorter continuous back-off duration, using an adaptive carrier sense threshold yields a negligible ($0\%$) occurrence of longer (greater than $3$~s) back-off duration, whereas there is a higher percentage of longer continuous back-off duration with static carrier sense thresholds. According to the IEEE $802.15.6$ BAN Standard guidelines, latency should be less than $125$ ms in medical applications and less than $250$ ms in non-medical applications \cite{tg6_std}. With the adaptive technique, the estimated average continuous back-off duration over the total time is $237$ ms ($<250$ ms). Also, the highest percentage (more than $7\%$ of the measured time) in Fig. \ref{bp} is for continuous back-off duration of $100$ ms ($<125$ ms) with this approach. On the other hand, with a static threshold of $-95$~dBm, channels can remain continuously in back-off over more than $1000$~s, almost $6\%$ of the time, and more importantly, the average continuous back-off duration within this threshold is up to $11$~s, which is very large for delay-constrained networks like BANs.

\vspace{-1mm}\subsection{Throughput (Successful Packets/s) vs. Packet Arrival Rate}
\begin{figure}[!t]
\centering{\includegraphics[width=\figwidth]{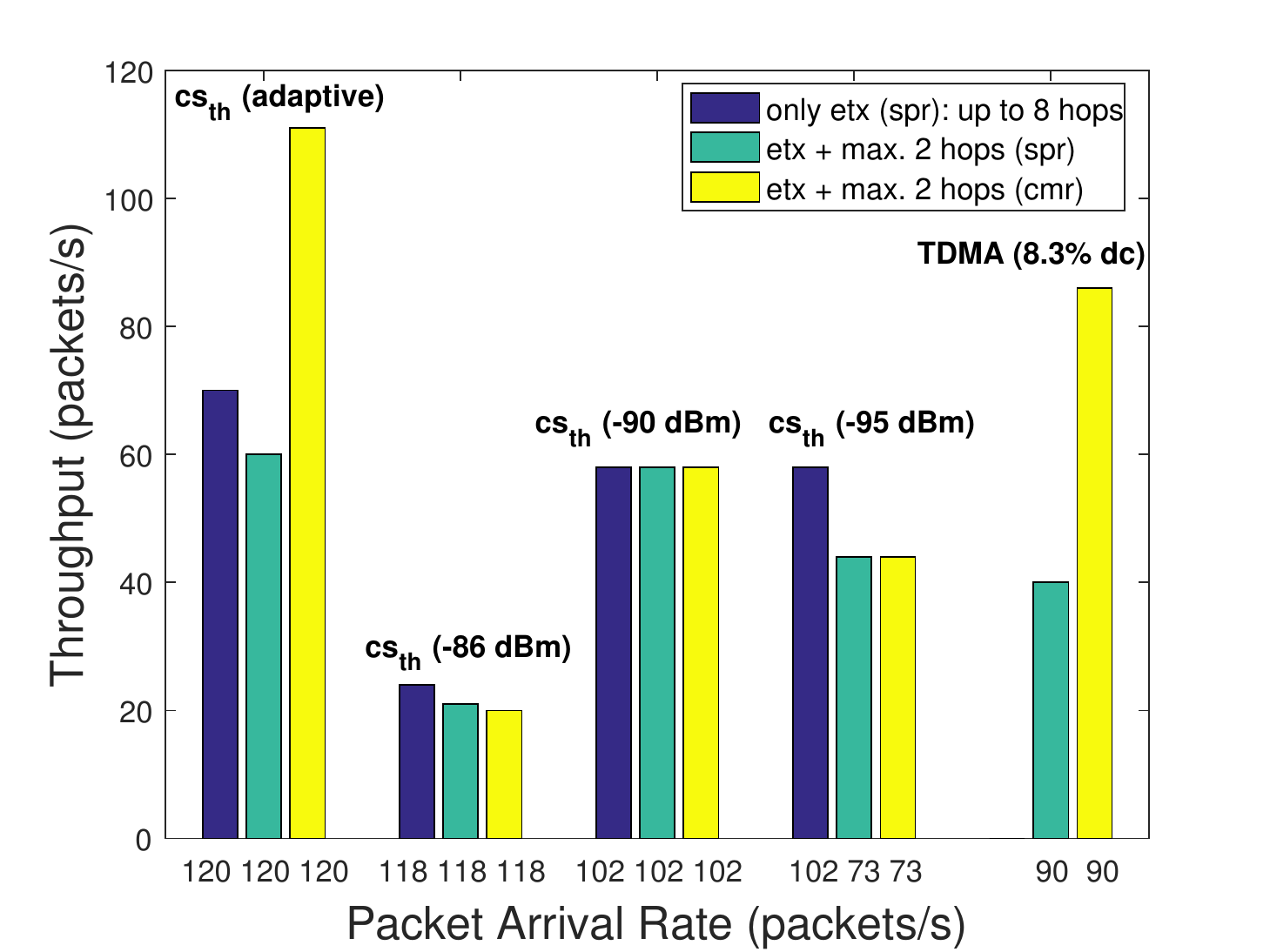}}
\caption{Average throughput (successful packets/s) vs. packet arrival rate over $10$ coexisting BANs for SPR and CMR (associated with CSMA/CA), with adaptive/static carrier sense thresholds ($CS_{th}$) and TDMA with $8.3\%$ duty cycle (dc). Transmit power $0$ dBm and receiver sensitivity $-90$ dBm}
\label{tvp}
\end{figure}

The average throughput (in terms of successful packets per-second) vs. the average packet arrival rate is shown in Fig. \ref{tvp}. It can be seen that the proposed adaptive CSMA/CA achieves the best result in terms of both throughput and packet arrival rate. Notably, with a higher static carrier sense threshold of $-86$ dBm the throughput is significantly lower with respect to packet arrival rate which indicates that, although more packets can be transmitted with a higher threshold (as it permits a higher interference level), the overall performance will degrade because of possibly lower SINR of the signal-of-interest. In Fig. \ref{tvp}, we also compare the CSMA/CA approach with a coordinated TDMA approach from \cite{shimly2017cross}, where the same setup is used with $4$ coordinated BANs receiving interference from $6$ non-coordinated nearby BANs. It is shown that adaptive CSMA/CA provides up to $20\%$ and $30\%$ improvement over TDMA (with a higher duty cycle of $8.3\%$), in terms of average throughput and packet arrival rate, respectively.
\setlength{\textfloatsep}{3pt}

\vspace{-1mm}
\subsection{Outage Probability with respect to SINR}
The outage probability with respect to SINR threshold can be expressed as,
\begin{equation}\label{outp}
P_{out} = Prob \big(\gamma_s < \gamma_{th}\big)
\end{equation}
where  $P_{out}$ is the probability of received SINR, $\gamma_s$, being less than a given threshold value $\gamma_{th}$. The signal-to-interference-plus-noise ratio (SINR) for any given link/branch is measured as follows:
\begin{equation}\label{sinr}
\gamma_s(\tau) = \frac{p_{tx} \left | h_{s,d}(\tau)\right | ^2}{\mathlarger{\sum}_{i=1}^{n} \big(p_{tx} \left | h_{int_i,d}(\tau)\right | ^2\big) + \left | \nu(\tau)\right | ^2}
\end{equation}
where $\gamma_s(\tau)$ is the measured SINR value of a signal $s$ at time instant $\tau$, $p_{tx}$ is the transmit power, $n$ is the number of interfering nodes, $\left | h_{s,d}\right |$ and $\left | h_{int_i,d}\right |$ are the average channel gains across the time instant of the signal-of-interest and the $i^{th}$ interfering signal, respectively. $\left | \nu \right |$ represents the instantaneous noise level at the destination node. The received noise power is set at $-100$ dBm.
\begin{figure}[!t]
\centering{\includegraphics[width=\figwidth]{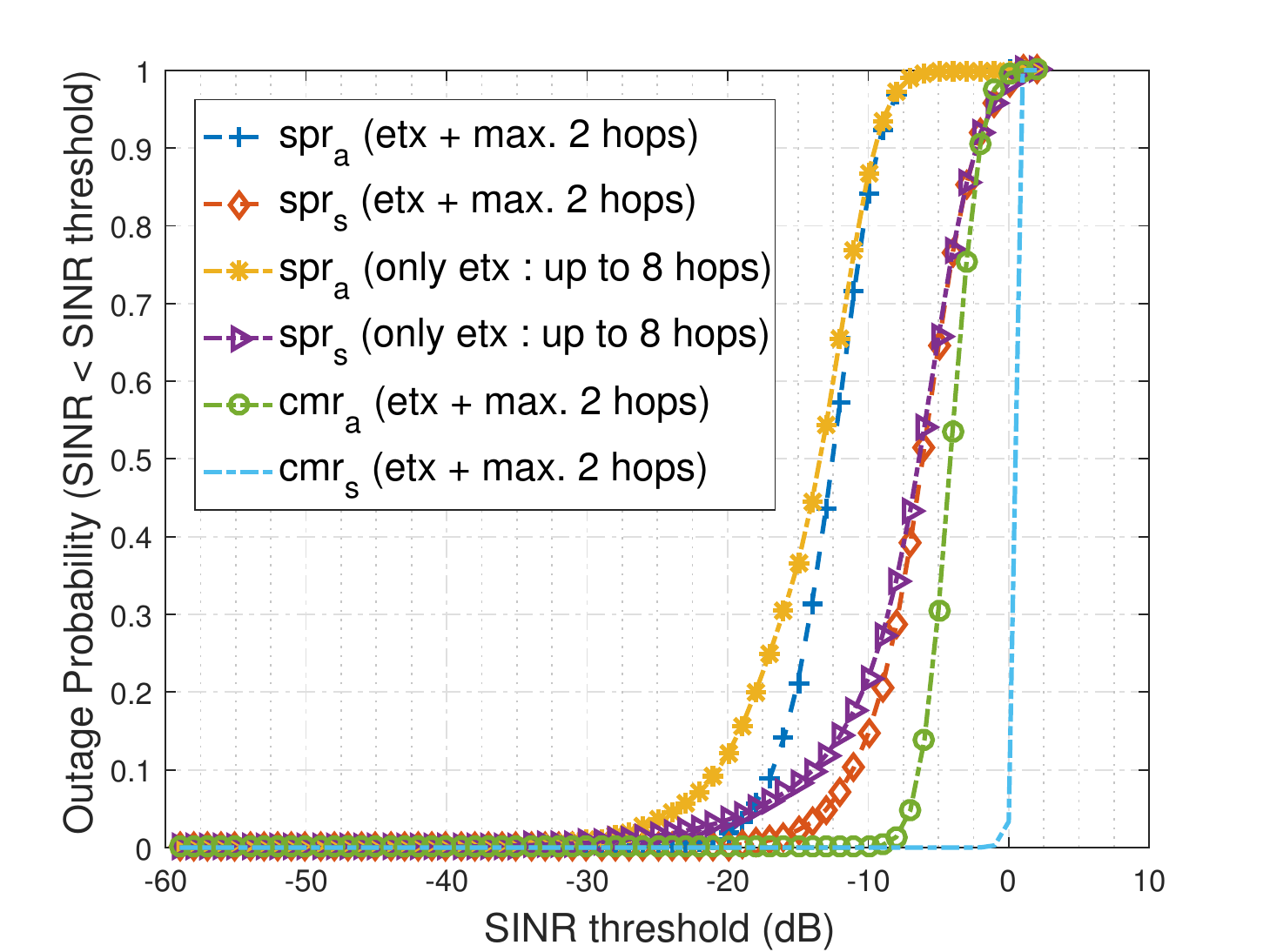}}
\caption{Average outage probability with respect to SINR thresholds for SPR and CMR (associated with CSMA/CA), with different routing metrics (e.g., only ETX, ETX + max. $2$ hops) for $10$ coexisting decentralized BANs; Subscript $a$ and $s$ refers to adaptive and static carrier sensing, respectively. Receiver sensitivity $-100$ dBm, transmit power $0$ dBm}
\label{sinr_fig}
\end{figure}
The averaged outage probability with respect to SINR thresholds found from SPR and CMR with adaptive and static ($-86$ dBm) carrier sensing CSMA/CA scheme applied over $10$ coexisting BANs is shown in Fig. \ref{sinr_fig}. It is demonstrated that, the use of a static threshold (i.e., $-86$ dBm) in CSMA/CA improves the outage probability with respect to SINR as it constantly avoids a significant interference level, although resulting in a longer back-off period (Fig. \ref{bp}) and throughput degradation (Fig. \ref{tvp}).

\subsection{Packet Delivery Ratio (PDR)}
\begin{figure}[!t]
\centering{\includegraphics[width=\figwidth]{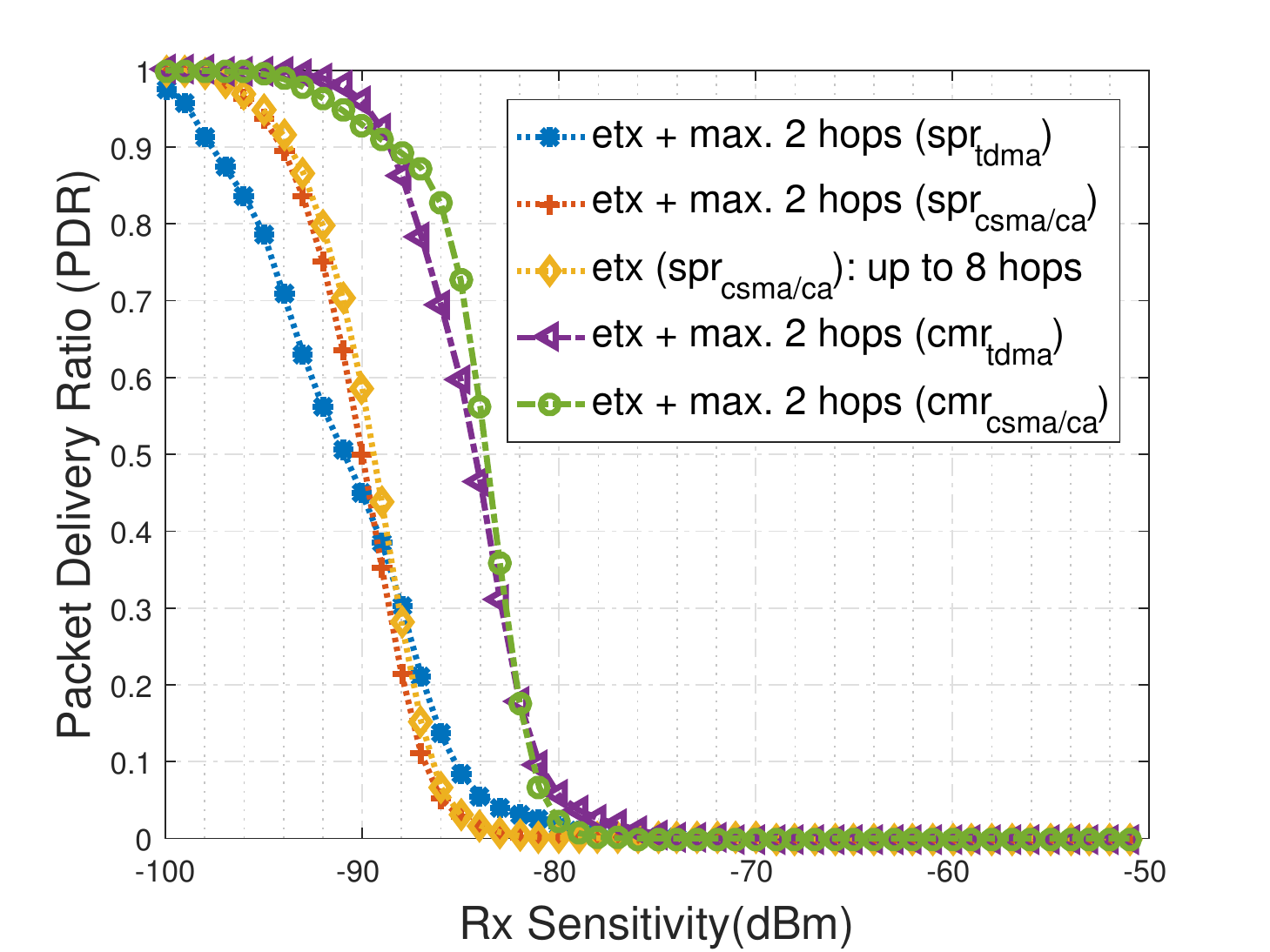}}
\caption{Average packet delivery ratio with respect to different receive sensitivities for SPR and CMR (associated with adaptive CSMA/CA and $8.3\%$ duty cycle TDMA), with different routing metrics (e.g., only ETX, ETX + max. $2$ hops) for $10$ coexisting BANs; Transmit power $0$ dBm}
\label{pdr}
\end{figure}

The averaged packet delivery ratio (PDR) with respect to different receive sensitivities for the given scheme with adaptive CSMA/CA and TDMA approach is presented in Fig. \ref{pdr}. It is shown that adaptive CSMA/CA yields better performance than a higher duty cycle TDMA. With a packet delivery ratio of $90\%$ (or packet error rate (PER) of $10\%$, as $PER = 1-PDR$), the SPR with adaptive CSMA/CA provides $4$ dB improvement over SPR with TDMA. It can also be seen that CMR gives more than $50\%$ (up to $65\%$) performance improvement over SPR with adaptive CSMA/CA, at $-88$ dBm receive sensitivity. Additionally, the best-case (at $-100$ dBm receive sensitivity) PDR for SPR and CMR is almost $100\%$ which is equivalent to a negligible PER (thus significantly surpassing the IEEE $802.15.6$ BAN Standard requirement of PER being less than $10\%$).

\subsection{Spectral Efficiency}
\begin{figure}[!t]
\centering{\includegraphics[width=\figwidth]{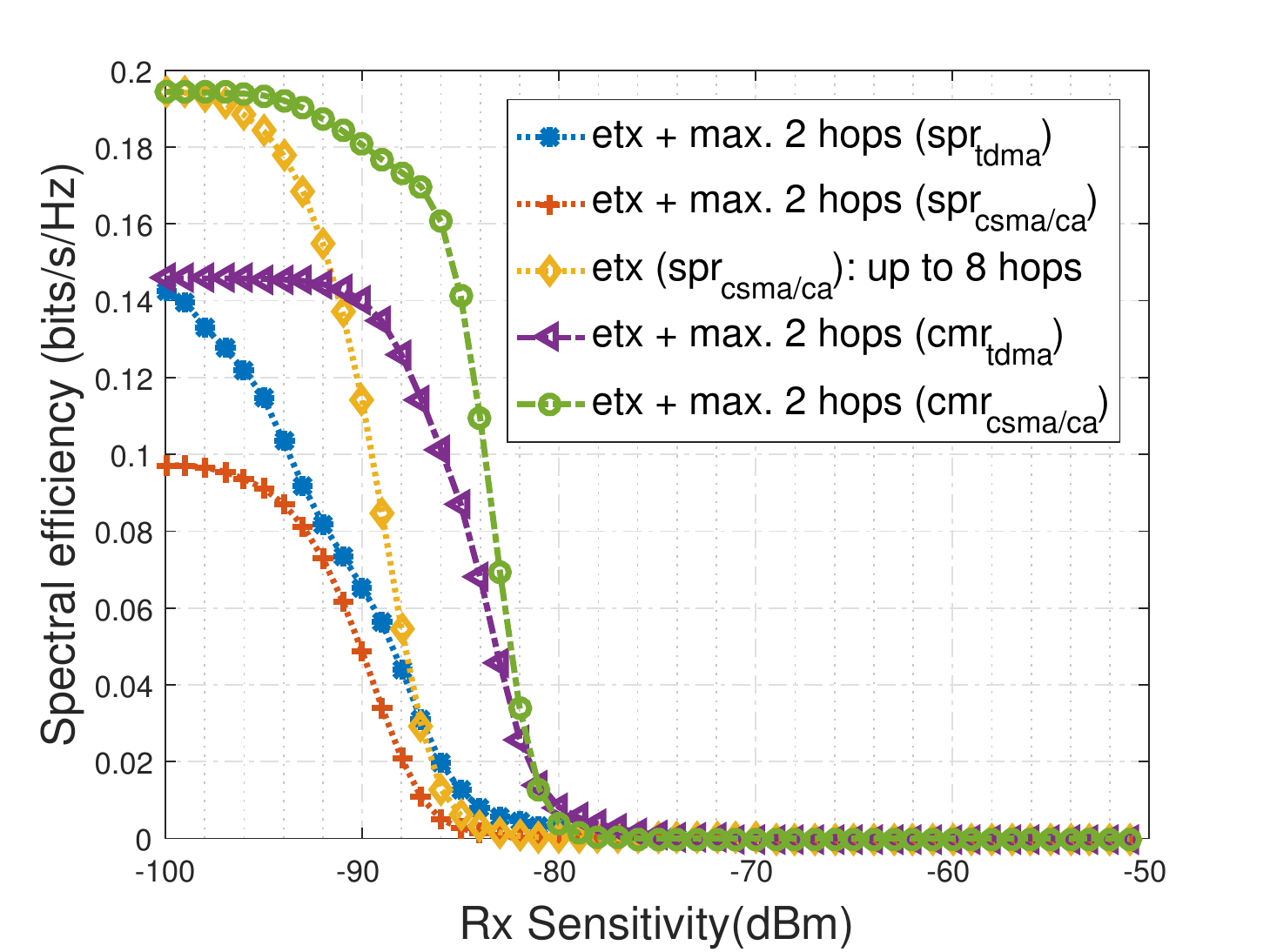}}
\caption{Average spectral efficiency for $10$ coexisting BANs with respect to different receive sensitivities for SPR and CMR (associated with adaptive CSMA/CA and $8.3\%$ duty cycle TDMA), with different routing metrics (e.g., only ETX, ETX + max. $2$ hops), at transmit power $0$ dBm}
\label{spef_fig}
\end{figure}

The spectral efficiency ($\zeta$) over BANs across which routing occurs is estimated as follows:
\begin{equation}\label{spef}
\zeta = \frac{\Theta \times n_c}{B}
\end{equation}
where $n_c$ is the number of actively routed BAN channels and $B$ is the bandwidth. The aggregated throughput can be defined as $(\Theta\times n_c)$, where the single channel throughput $\Theta$ can be measured as follows:
\begin{equation}\label{thr}
\Theta = \frac{P_{succ} \times \ell}{T}
\end{equation}
where $P_{succ}$ is the number of successfully delivered packets over the total time $T$ and $\ell$ is the length of the packet. The bandwidth and packet size can be found from Table \ref{table_param}, which are chosen in accordance with the IEEE $802.15.6$ Standard for narrowband communications \cite{tg6_std}. The average spectral efficiency of the overall network with respect to different receive sensitivities for adaptive CSMA/CA and higher ($8.3\%$) duty cycle TDMA are presented in Fig. \ref{spef_fig}. It can be seen that, adaptive CSMA/CA shows improvement over TDMA while contributing up to $6\%$ performance improvement over higher duty cycle TDMA at best-case scenario (at $-100$ dBm receive sensitivity). Even though, the adaptive CSMA/CA with SPR (ETX + max. $2$ hops) technique suffers from lower spectral efficiency because of the hop restriction, it shows improvement when performing SPR without any hop restriction due to the increased number of transmissions with longer paths.

\section{Conclusion}
We have proposed a simplified adaptive cross-layer carrier sensing mechanism for CSMA/CA in case of the coexistence of multiple wireless body area networks (BANs), which shows more than $50\%$ gain over static carrier sensing, in terms of throughput (successful packets/s) vs. packet arrival rate, as well as providing improved latency. We have also compared the performance of adaptive CSMA/CA with a coordinated TDMA approach by performing cross-layer optimized dynamic routing (i.e., shortest path routing, SPR, and cooperative multi-path routing, CMR), validated by experimental measurements. It is shown that adaptive CSMA/CA yields better performance than TDMA while providing up to $4$ dB, $20\%$ and $6\%$ improvement over higher ($8.3\%$) duty cycle TDMA in terms of PDR, throughput and spectral efficiency, respectively. The demonstrated feasibility of our method motivates its deployment in large-scale and highly-connected medical and non-medical applications. Planned future work includes applying Markov models for channel prediction, as well as combining adaptive TDMA and CSMA/CA to further improve BAN coexistence with higher throughput.


%

\bibliographystyle{IEEEtran}
\bibliography{Reference}

\end{document}